
\message
{JNL.TEX version 0.95 as of 5/13/90.  Using CM fonts.}

\catcode`@=11
\expandafter\ifx\csname inp@t\endcsname\relax\let\inp@t=\input
\def\input#1 {\expandafter\ifx\csname #1IsLoaded\endcsname\relax
\inp@t#1%
\expandafter\def\csname #1IsLoaded\endcsname{(#1 was previously loaded)}
\else\message{\csname #1IsLoaded\endcsname}\fi}\fi
\catcode`@=12

\font\twelverm=cmr12			\font\twelvei=cmmi12
\font\twelvesy=cmsy10 scaled 1200	\font\twelveex=cmex10 scaled 1200
\font\twelvebf=cmbx12			\font\twelvesl=cmsl12
\font\twelvett=cmtt12			\font\twelveit=cmti12
\font\twelvesc=cmcsc10 scaled 1200	\font\twelvesf=cmss12
                     
\font\twelvemib=cmmib10 scaled 1200
\font\tenmib=cmmib10
\font\eightmib=cmmib10 scaled 800

\skewchar\twelvei='177			\skewchar\twelvesy='60
\skewchar\twelvemib='177

\newfam\mibfam

\def\twelvepoint{\normalbaselineskip=12.4pt plus 0.1pt minus 0.1pt
  \abovedisplayskip 12.4pt plus 3pt minus 9pt
  \belowdisplayskip 12.4pt plus 3pt minus 9pt
  \abovedisplayshortskip 0pt plus 3pt
  \belowdisplayshortskip 7.2pt plus 3pt minus 4pt
  \smallskipamount=3.6pt plus1.2pt minus1.2pt
  \medskipamount=7.2pt plus2.4pt minus2.4pt
  \bigskipamount=14.4pt plus4.8pt minus4.8pt
  \def\rm{\fam0\twelverm}          \def\it{\fam\itfam\twelveit}%
  \def\sl{\fam\slfam\twelvesl}     \def\bf{\fam\bffam\twelvebf}%
  \def\mit{\fam 1}                 \def\cal{\fam 2}%
  \def\sc{\twelvesc}		   \def\tt{\twelvett}%
  \def\sf{\twelvesf}               \def\mib{\fam\mibfam\twelvemib}%
  \textfont0=\twelverm   \scriptfont0=\tenrm   \scriptscriptfont0=\sevenrm
  \textfont1=\twelvei    \scriptfont1=\teni    \scriptscriptfont1=\seveni
  \textfont2=\twelvesy   \scriptfont2=\tensy   \scriptscriptfont2=\sevensy
  \textfont3=\twelveex   \scriptfont3=\twelveex\scriptscriptfont3=\twelveex
  \textfont\itfam=\twelveit
  \textfont\slfam=\twelvesl
  \textfont\bffam=\twelvebf \scriptfont\bffam=\tenbf
                            \scriptscriptfont\bffam=\sevenbf
  \textfont\mibfam=\twelvemib \scriptfont\mibfam=\tenmib
                              \scriptscriptfont\mibfam=\eightmib
  \normalbaselines\rm}


\mathchardef\alpha="710B
\mathchardef\beta="710C
\mathchardef\gamma="710D
\mathchardef\delta="710E
\mathchardef\epsilon="710F
\mathchardef\zeta="7110
\mathchardef\eta="7111
\mathchardef\theta="7112
\mathchardef\iota="7113
\mathchardef\kappa="7114
\mathchardef\lambda="7115
\mathchardef\mu="7116
\mathchardef\nu="7117
\mathchardef\xi="7118
\mathchardef\pi="7119
\mathchardef\rho="711A
\mathchardef\sigma="711B
\mathchardef\tau="711C
\mathchardef\phi="711E
\mathchardef\chi="711F
\mathchardef\psi="7120
\mathchardef\omega="7121
\mathchardef\varepsilon="7122
\mathchardef\vartheta="7123
\mathchardef\varpi="7124
\mathchardef\varrho="7125
\mathchardef\varsigma="7126
\mathchardef\varphi="7127


\def\beginlinemode{\endmode
  \begingroup\parskip=0pt \obeylines\def\\{\par}\def\endmode{\par\endgroup}}
\def\beginparmode{\endmode
  \begingroup \def\endmode{\par\endgroup}}
\let\endmode=\par
{\obeylines\gdef\
{}}
\def\singlespace{\baselineskip=\normalbaselineskip}

\def\oneandahalfspace{\baselineskip=\normalbaselineskip
  \multiply\baselineskip by 3 \divide\baselineskip by 2}
\def\doublespace{\baselineskip=\normalbaselineskip \multiply\baselineskip by 2}

\newcount\firstpageno
\firstpageno=2
\footline={\ifnum\pageno<\firstpageno{\hfil}\else{\hfil\twelverm\folio\hfil}%
\fi}
\def\toppageno{\global\footline={\hfil}\global\headline
  ={\ifnum\pageno<\firstpageno{\hfil}\else{\hfil\twelverm\folio\hfil}\fi}}
\let\rawfootnote=\footnote		
\def\footnote#1#2{{\rm\singlespace\parindent=0pt\parskip=0pt
  \rawfootnote{#1}{#2\hfill\vrule height 0pt depth 6pt width 0pt}}}
\def\raggedcenter{\leftskip=4em plus 12em \rightskip=\leftskip
  \parindent=0pt \parfillskip=0pt \spaceskip=.3333em \xspaceskip=.5em
  \pretolerance=9999 \tolerance=9999
  \hyphenpenalty=9999 \exhyphenpenalty=9999 }
\def\dateline{\rightline{\ifcase\month\or
  January\or February\or March\or April\or May\or June\or
  July\or August\or September\or October\or November\or December\fi
  \space\number\year}}
\def\received{\vskip 3pt plus 0.2fill
 \centerline{\sl (Received\space\ifcase\month\or
  January\or February\or March\or April\or May\or June\or
  July\or August\or September\or October\or November\or December\fi
  \qquad, \number\year)}}


\hsize=6.5truein
\hoffset=0pt
\vsize=8.9truein
\voffset=0pt
\parskip=\medskipamount
\def\\{\cr}
\twelvepoint		
\doublespace		
\overfullrule=0pt	




\def\title			
  {\null\vskip 3pt plus 0.2fill
   \beginlinemode \doublespace \raggedcenter \bf}

\def\author			
  {\vskip 3pt plus 0.2fill \beginlinemode
   \singlespace \raggedcenter\sc}

\def\affil			
  {\vskip 3pt plus 0.1fill \beginlinemode
   \oneandahalfspace \raggedcenter \sl}

\def\abstract			
  {\vskip 3pt plus 0.3fill \beginparmode
   \oneandahalfspace ABSTRACT: }

\def\endtitlepage		
  {\endpage			
   \body}

\def\body			
  {\beginparmode}		

\def\head#1{			
  \goodbreak\vskip 0.5truein	
  {\immediate\write16{#1}
   \raggedcenter \uppercase{#1}\par}
   \nobreak\vskip 0.25truein\nobreak}

\def\itemitemitem{\par\indent\indent \hangindent3\parindent \textindent}
\def\itemitemitemitem{\par\indent\indent\indent \hangindent4\parindent
\textindent}
\def\beginitems{\par\medskip\bgroup
  \def\i##1 {\par\noindent\llap{##1\enspace}\ignorespaces}%
  \def\ii##1 {\item{##1}}%
  \def\iii##1 {\itemitem{##1}}%
  \def\iiii##1 {\itemitemitem{##1}}%
  \def\iiiii##1 {\itemitemitemitem{##1}}
  \leftskip=36pt\parskip=0pt}\def\enditems{\par\egroup}

\def\makefigure#1{\parindent=36pt\item{}Figure #1}

\def\figure#1 (#2) #3\par{\goodbreak\midinsert
\vskip#2
\bgroup\makefigure{#1} #3\par\egroup\endinsert}

\def\beneathrel#1\under#2{\mathrel{\mathop{#2}\limits_{#1}}}

\def\refto#1{$^{#1}$}		

\def\references			
  {\head{References}		
   \beginparmode
   \frenchspacing \parindent=0pt \leftskip=1truecm
   \parskip=8pt plus 3pt \everypar{\hangindent=\parindent}}

\gdef\refis#1{\item{#1.\ }}			

\gdef\journal#1, #2, #3, 1#4#5#6{		
    {\sl #1~}{\bf #2}, #3 (1#4#5#6)}		

\def\endreferences{\body}

\def\figurecaptions		
  {\endpage
   \beginparmode
   \head{Figure Captions}
}

\def\endpage			
  {\vfill\eject}

\def\endpaper			
  {\endmode\vfill\supereject}


\def\heading				
  {\vskip 0.5truein plus 0.1truein	
   \beginparmode \def\\{\par} \parskip=0pt \singlespace \raggedcenter}

\def\subheading				
  {\vskip 0.25truein plus 0.1truein	
   \beginlinemode \singlespace \parskip=0pt \def\\{\par}\raggedcenter}

\def\tag#1$${\eqno(#1)$$}

\def\align#1$${\eqalign{#1}$$}

\def\aligntag#1$${\gdef\tag##1\\{&(##1)\cr}\eqalignno{#1\\}$$
  \gdef\tag##1$${\eqno(##1)$$}}

\def\endaligntag{}

\def\overset #1\to#2{{\mathop{#2}\limits^{#1}}}
\def\underset#1\to#2{{\let\next=#1\mathpalette\undersetpalette#2}}
\def\undersetpalette#1#2{\vtop{\baselineskip0pt
\ialign{$\mathsurround=0pt #1\hfil##\hfil$\crcr#2\crcr\next\crcr}}}


\def\ref#1{Ref.~#1}			
\def\Ref#1{Ref.~#1}			
\def\[#1]{[\cite{#1}]}
\def\cite#1{{#1}}
\def\(#1){(\call{#1})}
\def\call#1{{#1}}
\def\taghead#1{}
\def\frac#1#2{{#1 \over #2}}

\def\12{{1\over2}}

\def\sla{\raise.15ex\hbox{$/$}\kern-.57em}
\def\leaderfill{\leaders\hbox to 1em{\hss.\hss}\hfill}
\def\twiddle{\lower.9ex\rlap{$\kern-.1em\scriptstyle\sim$}}
\def\bigtwiddle{\lower1.ex\rlap{$\sim$}}
\def\gtwid{\mathrel{\raise.3ex\hbox{$>$\kern-.75em\lower1ex\hbox{$\sim$}}}}
\def\ltwid{\mathrel{\raise.3ex\hbox{$<$\kern-.75em\lower1ex\hbox{$\sim$}}}}
\def\square{\kern1pt\vbox{\hrule height 1.2pt\hbox{\vrule width 1.2pt\hskip 3pt
   \vbox{\vskip 6pt}\hskip 3pt\vrule width 0.6pt}\hrule height 0.6pt}\kern1pt}
\def\tdot#1{\mathord{\mathop{#1}\limits^{\kern2pt\ldots}}}
\def\happyface{%
$\bigcirc\rlap{\lower0.3ex\hbox{$\kern-0.85em\scriptscriptstyle\smile$}%
\raise0.4ex\hbox{$\kern-0.6em\scriptstyle\cdot\cdot$}}$}
\def\sadface{%
$\bigcirc\rlap{\lower0.25ex\hbox{$\kern-0.85em\scriptscriptstyle\frown$}%
\raise0.43ex\hbox{$\kern-0.6em\scriptstyle\cdot\cdot$}}$}

\def\pmb#1{\setbox0=\hbox{#1}%
  \kern-.025em\copy0\kern-\wd0
  \kern  .05em\copy0\kern-\wd0
  \kern-.025em\raise.0433em\box0 }

\catcode`@=11
\newcount\r@fcount \r@fcount=0
\newcount\r@fcurr
\immediate\newwrite\reffile
\newif\ifr@ffile\r@ffilefalse
\def\w@rnwrite#1{\ifr@ffile\immediate\write\reffile{#1}\fi\message{#1}}

\def\writer@f#1>>{}
\def\referencefile{
  \r@ffiletrue\immediate\openout\reffile=\jobname.ref%
  \def\writer@f##1>>{\ifr@ffile\immediate\write\reffile%
    {\noexpand\refis{##1} = \csname r@fnum##1\endcsname = %
     \expandafter\expandafter\expandafter\strip@t\expandafter%
     \meaning\csname r@ftext\csname r@fnum##1\endcsname\endcsname}\fi}%
  \def\strip@t##1>>{}}

\def\citeall#1{\xdef#1##1{#1{\noexpand\cite{##1}}}}
\def\cite#1{\each@rg\citer@nge{#1}}	

\def\each@rg#1#2{{\let\thecsname=#1\expandafter\first@rg#2,\end,}}
\def\first@rg#1,{\thecsname{#1}\apply@rg}	
\def\apply@rg#1,{\ifx\end#1\let\next=\relax
\else,\thecsname{#1}\let\next=\apply@rg\fi\next}

\def\citer@nge#1{\citedor@nge#1-\end-}	
\def\citer@ngeat#1\end-{#1}
\def\citedor@nge#1-#2-{\ifx\end#2\r@featspace#1 
  \else\citel@@p{#1}{#2}\citer@ngeat\fi}	
\def\citel@@p#1#2{\ifnum#1>#2{\errmessage{Reference range #1-#2\space is bad.}%
    \errhelp{If you cite a series of references by the notation M-N, then M and
    N must be integers, and N must be greater than or equal to M.}}\else%
 {\count0=#1\count1=#2\advance\count1
by1\relax\expandafter\r@fcite\the\count0,%
  \loop\advance\count0 by1\relax
    \ifnum\count0<\count1,\expandafter\r@fcite\the\count0,%
  \repeat}\fi}

\def\r@featspace#1#2 {\r@fcite#1#2,}	
\def\r@fcite#1,{\ifuncit@d{#1}
    \newr@f{#1}%
    \expandafter\gdef\csname r@ftext\number\r@fcount\endcsname%
                     {\message{Reference #1 to be supplied.}%
                      \writer@f#1>>#1 to be supplied.\par}%
 \fi%
 \csname r@fnum#1\endcsname}
\def\ifuncit@d#1{\expandafter\ifx\csname r@fnum#1\endcsname\relax}%
\def\newr@f#1{\global\advance\r@fcount by1%
    \expandafter\xdef\csname r@fnum#1\endcsname{\number\r@fcount}}

\let\r@fis=\refis			
\def\refis#1#2#3\par{\ifuncit@d{#1}
   \newr@f{#1}%
   \w@rnwrite{Reference #1=\number\r@fcount\space is not cited up to now.}\fi%
  \expandafter\gdef\csname r@ftext\csname r@fnum#1\endcsname\endcsname%
  {\writer@f#1>>#2#3\par}}

\def\ignoreuncited{
   \def\refis##1##2##3\par{\ifuncit@d{##1}%
     \else\expandafter\gdef\csname r@ftext\csname
r@fnum##1\endcsname\endcsname%
     {\writer@f##1>>##2##3\par}\fi}}

\def\r@ferr{\endreferences\errmessage{I was expecting to see
\noexpand\endreferences before now;  I have inserted it here.}}
\let\r@ferences=\references
\def\references{\r@ferences\def\endmode{\r@ferr\par\endgroup}}

\let\endr@ferences=\endreferences
\def\endreferences{\r@fcurr=0
  {\loop\ifnum\r@fcurr<\r@fcount
    \advance\r@fcurr by 1\relax\expandafter\r@fis\expandafter{\number\r@fcurr}%
    \csname r@ftext\number\r@fcurr\endcsname%
  \repeat}\gdef\r@ferr{}\endr@ferences}


\let\r@fend=\endpaper\gdef\endpaper{\ifr@ffile
\immediate\write16{Cross References written on []\jobname.REF.}\fi\r@fend}

\catcode`@=12

\citeall\refto		
\citeall\ref		%
\citeall\Ref		%

\catcode`@=11
\newcount\tagnumber\tagnumber=0

\immediate\newwrite\eqnfile
\newif\if@qnfile\@qnfilefalse
\def\write@qn#1{}
\def\writenew@qn#1{}
\def\w@rnwrite#1{\write@qn{#1}\message{#1}}
\def\@rrwrite#1{\write@qn{#1}\errmessage{#1}}

\def\taghead#1{\gdef\t@ghead{#1}\global\tagnumber=0}
\def\t@ghead{}

\expandafter\def\csname @qnnum-3\endcsname
  {{\t@ghead\advance\tagnumber by -3\relax\number\tagnumber}}
\expandafter\def\csname @qnnum-2\endcsname
  {{\t@ghead\advance\tagnumber by -2\relax\number\tagnumber}}
\expandafter\def\csname @qnnum-1\endcsname
  {{\t@ghead\advance\tagnumber by -1\relax\number\tagnumber}}
\expandafter\def\csname @qnnum0\endcsname
  {\t@ghead\number\tagnumber}
\expandafter\def\csname @qnnum+1\endcsname
  {{\t@ghead\advance\tagnumber by 1\relax\number\tagnumber}}
\expandafter\def\csname @qnnum+2\endcsname
  {{\t@ghead\advance\tagnumber by 2\relax\number\tagnumber}}
\expandafter\def\csname @qnnum+3\endcsname
  {{\t@ghead\advance\tagnumber by 3\relax\number\tagnumber}}

\def\equationfile{%
  \@qnfiletrue\immediate\openout\eqnfile=\jobname.eqn%
  \def\write@qn##1{\if@qnfile\immediate\write\eqnfile{##1}\fi}
  \def\writenew@qn##1{\if@qnfile\immediate\write\eqnfile
    {\noexpand\tag{##1} = (\t@ghead\number\tagnumber)}\fi}
}

\def\callall#1{\xdef#1##1{#1{\noexpand\call{##1}}}}
\def\call#1{\each@rg\callr@nge{#1}}

\def\each@rg#1#2{{\let\thecsname=#1\expandafter\first@rg#2,\end,}}
\def\first@rg#1,{\thecsname{#1}\apply@rg}
\def\apply@rg#1,{\ifx\end#1\let\next=\relax%
\else,\thecsname{#1}\let\next=\apply@rg\fi\next}

\def\callr@nge#1{\calldor@nge#1-\end-}
\def\callr@ngeat#1\end-{#1}
\def\calldor@nge#1-#2-{\ifx\end#2\@qneatspace#1 %
  \else\calll@@p{#1}{#2}\callr@ngeat\fi}
\def\calll@@p#1#2{\ifnum#1>#2{\@rrwrite{Equation range #1-#2\space is bad.}
\errhelp{If you call a series of equations by the notation M-N, then M and
N must be integers, and N must be greater than or equal to M.}}\else%
 {\count0=#1\count1=#2\advance\count1
by1\relax\expandafter\@qncall\the\count0,%
  \loop\advance\count0 by1\relax%
    \ifnum\count0<\count1,\expandafter\@qncall\the\count0,%
  \repeat}\fi}

\def\@qneatspace#1#2 {\@qncall#1#2,}
\def\@qncall#1,{\ifunc@lled{#1}{\def\next{#1}\ifx\next\empty\else
  \w@rnwrite{Equation number \noexpand\(>>#1<<) has not been defined yet.}
  >>#1<<\fi}\else\csname @qnnum#1\endcsname\fi}

\let\eqnono=\eqno
\def\eqno(#1){\tag#1}
\def\tag#1$${\eqnono(\displayt@g#1 )$$}

\def\aligntag#1\endaligntag
  $${\gdef\tag##1\\{&(##1 )\cr}\eqalignno{#1\\}$$
  \gdef\tag##1$${\eqnono(\displayt@g##1 )$$}}

\def\eqalignno#1{\displ@y \tabskip\centering
  \halign to\displaywidth{\hfil$\displaystyle{##}$\tabskip\z@skip
    &$\displaystyle{{}##}$\hfil\tabskip\centering
    &\llap{$\displayt@gpar##$}\tabskip\z@skip\crcr
    #1\crcr}}

\def\displayt@gpar(#1){(\displayt@g#1 )}

\def\displayt@g#1 {\rm\ifunc@lled{#1}\global\advance\tagnumber by1
        {\def\next{#1}\ifx\next\empty\else\expandafter
        \xdef\csname @qnnum#1\endcsname{\t@ghead\number\tagnumber}\fi}%
  \writenew@qn{#1}\t@ghead\number\tagnumber\else
        {\edef\next{\t@ghead\number\tagnumber}%
        \expandafter\ifx\csname @qnnum#1\endcsname\next\else
        \w@rnwrite{Equation \noexpand\tag{#1} is a duplicate number.}\fi}%
  \csname @qnnum#1\endcsname\fi}

\def\ifunc@lled#1{\expandafter\ifx\csname @qnnum#1\endcsname\relax}

\let\@qnend=\end\gdef\end{\if@qnfile
\immediate\write16{Equation numbers written on []\jobname.EQN.}\fi\@qnend}

\catcode`@=12

\def\Had{ H_{A\bar \gamma}}
\rightline{NSF-ITP-92-103}
\rightline{TPI-MINN-92-33/T}
\rightline{June  1992}
\vskip.8in
\centerline{\bf
The Rule of Discarding $1/N_c$ in Inclusive Weak Decays
}
\bigskip
\centerline{B. Blok
}
\bigskip
\centerline{\sl Institute for Theoretical Physics}
\centerline{\sl University of California at Santa Barbara}
\centerline{\sl Santa Barbara, CA 93106 }
\centerline{\sl and}
\centerline{M. Shifman
}
\bigskip
\centerline{\sl  Theoretical Physics Institute}
\centerline{\sl University of Minnesota}
\centerline{\sl Minnesota, MN 55455}
\bigskip
\abs{ We continue to investigate
the effects of the (dominant) subleading
 operator of the $\vec\sigma\vec H$
type in the $1/N_c$ parts of the weak
 nonleptonic amplitudes. If the previous
work concentrated on exclusive decays now we analyse inclusive widths in a
certain kinematic limit. Deviations from the parton model predictions are
found due to soft-gluon color exchange. In the interference (i.e. $1/N_c$)
terms they are stronger than in the non-interference terms. The sign is
negative so that the $1/N_c$ parts tend
 to cancel. The absolute value of the
effect is of order 1 in D-mesons and of order 0.50 in B mesons. }
\endpage
\head{1. Introduction}
\par In our previous work \refto{1}
 a QCD-based method is proposed allowing
 one to get model independent estimates of the nonfactorizable
$1/N_c$ parts of
weak nonleptonic amplitudes. The analysis
 of \ref{1} is limited to exclusive
transitions. It turns out, however, that the very same ideas ensure a
 significant breakthrough in understanding
  of inclusive decays. A number of
 puzzles known in the literature for years
  acquire a satisfactory solution at
a semiquantative level, and a very appealing general qualitative picture
emerges.
\par As well known, inclusive nonleptonic widths of hadrons with one
heavy quark in the limit $m_Q\rightarrow \infty$ are given by the parton
model prescribing to calculate
 the width of a free decay of a heavy quark into
three light quarks. (Logarithms due to hard gluon exchange should be, of
course, taken into account in the effective four-fermion
 Lagrangian $L^{W}_{\rm eff}$).
  The actual heavy quarks have finite masses, however,
and a reach dynamical pattern of inclusive widths, at least in the case of
the c quarks, which  shows that preasymptotic corrections are quite
substantial.
\par Theoretical
QCD-based analysis of various preasymptotic effects due to the
spectator quarks has been carried out in \ref{2,3,4} . The most important
 preasymptotic mechanism revealed in the analysis of \ref{2,3,4} is the
interference mechanism suppressing the $D^+$ width by a factor of
$\sim 2$
respectively to that of $D^0$.
 This suppression is welcome phenomenologically.
\par At the same time, the semileptonic branching ratio of $D^+$ is close
to the prediction of the
 parton model, while the semileptonic branching ratio
of $D^0$, ${\rm Br}(D^0\rightarrow e^+ +{\rm anything})$,
 looks suppressed by a factor of $\sim 2$. Thus,
 the analysis of \ref{2,3,4} had, obviously, a missing element which was
badly needed: an overall
 enhancement of the c-quark decay rate, not related to
the spectator quarks.
\par A seemingly unrelated
 evolution line in the theory of weak nonleptonic
decays is the so-called rule of discarding $1/N_c$. The observation of the
failure of the naive
 factorization technique (for a detailed description of
this technique see \ref{5} )
 was first made in \ref{6}. It was then formulated
as a general empiric rule of of discarding $1/N_c$ parts of the
amplitudes. \refto{7} Hints on dynamical mechanism which might
explain this rule are known
 in literature \refto{8,9,10,22,11}. Of most importance
 for the present discussion is the recent idea of
the color exchange due to a soft gluon reducible to a
magnetic type operator $\vec \sigma \vec H$ \refto{1} (see below).
\par A conjecture that the interference ( e.g. $1/N_c$) parts can be
 dynamically suppressed not
  only in exclusive channels but in the inclusive
 rate as well $-$ and if so,
  this would universally enhance all charmed meson
and baryon decays and resolve
 the above paradox $-$ is explicitly formulated in
\ref{7} (see also
\ref{5} ). Moreover, it was  argued \refto{11} that if we move
from the point $m_Q=\infty$ to finite heavy quark masses the preasymptotic
 corrections start earlier in the interference terms than they start
in noninterference terms. As a result the parton model predictions are
stronger distorted (suppressed) in $1/N_c$ parts.
\par Below we will demonstrate using the approach of \ref{1}
that both conjectures do actually take place in QCD.
\par It is worth noting that
 all preasymptotic effects discussed previously
\refto{2,3,4} have the relative weight $\sim f^2_Q/m_Q^2$, where $f_Q $
is a pseudoscalar meson coupling constant (analog of $f_\pi$) and
$m_Q$ is the heavy quark mass. Since asymptotically $f^2_Q\sim m^{-1}_Q$
\refto{12} these corrections scale as $m^{-3}_Q$. The existence of the
preasymptotic effects scaling like $m^{-2}_Q$ is observed in \ref{13}
devoted in part to the same question as the present work,
with a considerable overlap of results (\ref{13} appeared when
our paper was in preparation).
 Therefore, theoretically, the corrections proportional to $m_Q^{-2}$
   are the leading power corrections which happened
 to be connected with the operator
 $\bar Q\Gamma^\mu t^ag\tilde G^a_{\alpha\mu}Q$,
playing a special role
 in the color exchange and in the interference terms in
general. Numerically, though, for
 the charmed quark they are not more important
than the interference corrections of
 \ref{2,3,4}. For the b quark the $m_Q^{-2}$
law wins over $m^{-3}_Q$, and the
 $m^{-2}_Q$ correction seems to dominate the
deviations from the parton model.
\head{2. Interference terms and their role in the asymptotic regime.}
\par We open this section with a
 remark on kinematics. Whatever the mass of the
heavy quark is, we assume
that the  W boson mass is much
 larger, so that all weak nonleptonic amplitudes
can be described by an effective
 four-fermion Lagrangian. The coefficients of
this Lagrangian carry information on the structure of flavordynamics
(CKM matrix elements) and, also, they take into account short distance
physics (distances from $m_Q^{-1}$ to $ m^{-1}_W$).
 In particular, the effects due to hard gluon
exchanges are included. We will disregard the penguin graphs \refto{14}
which are irrelevant in this range of questions. Then the effective weak
 hamiltonian has the generic form
$$\L_{\rm eff}={G_F\over \sqrt{2}}V_1V_2(c_1(\bar B\Gamma^\mu A)
(\bar \beta\Gamma_\mu\alpha)+c_2(\bar B^i\Gamma^\mu A_j)
(\bar \beta^j\Gamma_\mu\alpha_i))\eqno (1)$$
where $G_F$ is the Fermi coupling constant, $V_{1,2}$ are the relevant CKM
matrix elements, $c_{1,2}$ are numerical coefficients, $\Gamma^\mu=
\gamma^\mu(1+\gamma^5)$, A is
 the initial heavy quark, $\alpha$ and $\beta$
are the final massless quarks,
 B is the final  quark whose mass will be
treated as a free parameter. We consider $V_{1,2}$ and $c_{1,2}$ as given
numbers and do not discuss them at all. The coefficients $c_{1,2}$ were
calculated long ago \refto{15} (for a
 pedagogical discussion see \ref{16}).
The focus of our analysis is the conversion of the quark Lagrangian
\(1) in the actual hadronic transition.
 As for applications, a few practical
examples are:
\par (i) c$\rightarrow$ su$\bar d$ (A=c, B=s, $\alpha =d$,
 $\beta =u$);
\par (ii) $b\rightarrow cd \bar u$ (A=b,
 B=c, $\alpha =u$, $\beta =d$);
\par (iii) Cabibbo suppressed $b\rightarrow ud\bar u$ (A=b, B=u,
$\alpha =u, \beta =d$).
\par In our previous work \refto{1} we
 discussed the exclusive processes, and
the method suggested for treating soft
 color exchange  is applicable
in the kinematical domain
$$m_{A,B}\rightarrow \infty, m_A-m_B=\Delta\sim GeV\eqno (2)$$
where $\Delta$ is held fixed and is of order of a typical hadronic scale.
Now we pass to the discussion of the
 inclusive rates based on the asymptotic
 formulae. Correspondingly, we can not
  work in the kinematic region accepted
in \ref{1} . Our present analyses is parametrically justified in the limit
$$m_A\rightarrow \infty , \Delta=m_A-m_B\gg \Lambda_{\rm QCD}.\eqno (3)$$
B quark may be either massless or heavy; the mass difference $\Delta$
may or may not scale with $m_A$-this is unimportant. What is important,
this parameter is assumed to be much
 larger than the typical hadronic scale,
so that we are able to consider expansion in $\Delta^{-1}$.
\par In practice, there exists a boundary domain in which it is possible
to use, at least semiquantatively, both languages: that based on
eq. \(2) and on eq. \(3).  The actual decays (i) and (ii) are of this
type. So, one may hope that the method \refto{1} and considerations presented
in this paper give reasonable estimates for the soft color exchange
both in the exclusive and inclusive transitions in these cases. Further
comments on matching the domains \(2) and \(3) are given below.
At first, we focus on the analysis of the inclusive decays in the
kinematics \(3) .
\par The inclusive widths of the
 A-flavored hadrons stemming from Lagrangian
\(1) receive two distinct contributions. The leading in $N_c$ (the number
of colors) contribution comes from the squares of the matrix elements
of the
 first and the second terms. It is proportional
  to $N_c(\vert c_1\vert^2+\vert c_2\vert^2)$.
If we treat $N_c$ as a large parameter and keep only the leading and
subleading terms in $1/N_c$ the calculation of $c_1^2$ and $c_2^2$ parts
basically factorizes in two blocks (Fig. 1).
\par Each block includes all sorts of QCD corrections, but there is no
 communication between them (the
  latter appears only at the level $1/N^2_c$).
Therefore the question of validity of the parton
  results is decided for each
block separately. For instance, the $c_1^2$ part is the product of the
 polarization operator
induced by the currents $\bar\beta\Gamma^\mu\alpha$
 times the square of an amplitude which is
effectively the same as in the semileptonic $A\rightarrow B$ transition.
At large $m_A$ (more exactly, $\Delta$) both
 amplitudes are given by the parton
model. The asymptotic (power) corrections to both amplitudes are of order
$m_A^{-4}$ ($\Delta^{-4}$).
\footnote*{Let us emphasise once more that we consider here only those
effects which do not depend on the flavor of the spectator quark.}
 Thus, they are
 parametrically smaller than those
emerging in the interference term (see below).
\par Neglecting all power corrections we get for $\Gamma
 (A\rightarrow B\bar \alpha
\beta )$ in the asymptotic regime:
$$\Gamma ={ G_F^2\vert V_1V_2\vert^2N_c\over
 192\pi^3}m_A^5F_1({m^2_B\over m^2_A})
(c_1^2+c_2^2+2c_1c_2/N_c).\eqno (4)$$
Here the function $F_1$ takes into account the effect of the nonvanishing
 B-quark mass; it is normalized in such a
  way that $F_1(0)=1$. Calculation of
this function is well known in literature, e. g.
 \ref{17} (a simple derivation is
sketched in the Appendix):
$$F_1(x)=(1-x^2)(1-8x+x^2)-12x^2{\rm ln}(x).\eqno (4a)$$
\par The interference term suppressed
 by $1/N_c$ is destructive due to the fact
that the coefficients $c_1$ and $c_2$ have opposite signs. Specifically
 \refto{16} :
$$\eqalign{&c_1\sim 1.35\quad c_2\sim -0.52\quad {\rm for}\quad c
\rightarrow s\bar
 u\bar d,\cr
&c_1\sim 1.26\quad c_2\sim -0.29\quad
 {\rm for}\quad b\rightarrow c\bar u d.\cr}
\eqno  (5)$$
Therefore, if the leading order preasymptotic
  corrections do indeed occur
at first \footnote{**}{By saying "at first" we mean that we move from the
asymptotic point $m_A=\infty $ towards smaller values of quark mass.}
in the interference term tending to cancel it, this will mean that the  total
widths are universally enhanced $-$ the tendency welcomed from at least two
 points of view. It solves the difficulty with $\Gamma (D^0)/\Gamma (D^+)$
(suppression of $\Gamma (D^+)$ by the  mechanism \refto{2,3,4} with
seemingly undistorted
semileptonic width ${\rm \Gamma}_{\rm sl}(D^+))$
 and helps understand a somewhat
smaller experimental value of
 ${\rm \Gamma}_{\rm sl}(B)$ than
   expected in the naive partonic model.
Of course, this result crucially
 depends on numericals. We will discuss the
situation in detail in the next section.
\par Here we would like to make a
 remark
referring to the asymptotic regime
\(4), where the exact quark hadron duality takes place.
\par For the non-interference terms $\sim (c_1^2+c_2^2)$ it is easy to see
that the standard duality prescription is perfectly compatible with the
standard factorization of the graph
 of Fig.1 into two blocks. The duality is
implemented in each block separately and saturating by the hadronic states
we get precisely the $c_1^2+c_2^2$
 part of eq. \(4), at least , at the level
 of the leading terms in $N_c$ and the next-to-leading terms. The issue of
saturation and duality becomes less
 trivial  only at the level of $1/N^2_c$
where the two factorized blocks start
 interacting. At this level the problem
becomes nontrivial, but we will not pursue this issue further here.
\par The situation with the interference
 term $\sim c_1c_2$ is quite different.
Here the naive factorization prescription
 is incompatible with duality from
the very beginning (let us parenthetically note that the use of naive
factorization
is a common practice in weak nonleptonic decays, although theoretically
it is perhaps justified only in exceptional kinematical circumstances
, see \ref{18} and discussion below.)
\par Indeed, at the quark level the interference term
for the decay of $A$-flavored heavy hadron $\Had$ is given by
($\bar \gamma$ is the spectator light antiquark)
$$\eqalign{&\Delta\Gamma_{\rm naive}(\Had )={G^2_F\over 2}
\vert V_1V_2\vert^2 2c_1c_2{1\over m_{\Had}} {\rm Im} <\Had\vert\int d^4x
iT\{ O_1(x)O_2^+(0)\}\vert\Had >,\cr
&O_1=(\bar B\Gamma_\mu A)(\bar \beta \Gamma_\mu \alpha );\,\,\,\,
O_2=(\bar B^i\Gamma_\mu A_j)(\bar \beta^j
\Gamma_\mu \alpha_i).\cr}\eqno (6)$$
After simple calculation we get
$$\Delta\Gamma_{\rm naive}=\vert V_1V_2\vert^2c_1c_2G^2_F
 {m_A^5\over 96\pi^3}F_1 (m_B^2/m_A^2).\eqno (7)$$
Let us try to understand
 how one could saturate this result by the hadronic
states within the naive factorization.
\par Since the intermediate hadronic states should have nonvanishing
projection both on $O_1$ and $O_2$, there are three possible options
 corresponding to three different assignments of color factors ( we remind
that $1/N_c$ is formally considered as a large parameter):
\par (i) $O_1$ produces a
 hadronic state at the leading order; the projection
of this state on $O_2$ contains $1/N_c$ factor.
\par (ii) An intermediate hadronic
 state is produced by $O_2$ at the leading
order with the projection on $O_1$ suppressed by $1/N_c$
\par (iii) An intermediate hadronic
 state has both projections on $O_1$ and
 $O_2$ suppressed by ${1\over \sqrt{N_c}}$.

An example of the state of the type
 (i) is $(\beta\bar\alpha)+(B\bar\gamma )$
where $\bar \gamma$ is the spectator antiquark composing the initial
 $\Had $, the letters in
  brackets denote the quark content of the
meson considered.\footnote*{In each
 channel ground states as well as possible
excitations should be included in the spirit of duality}
The state of the type (ii) can be written as
 $(B\bar\alpha )+(\beta
 \bar\gamma)$. Finally,the intermediate states of the
 type (iii) must contain an
 extra quark-antiquark meson, and they are of the form
$$\eqalign{&(B\bar\beta)+(\beta\bar\gamma )+(\beta\bar\alpha )\quad
-{\rm extra}\quad \bar\beta\beta \quad{\rm pair}\cr
&{\rm or}\cr
&(B\bar\gamma )+(B\bar\alpha )+(\beta\bar B )\quad
-{\rm extra}\quad \bar BB \quad{\rm pair}.\cr}$$
The second variant can
 be disregarded because of the mass of B-quark and the
corresponding suppression due to the production of the massive pair
$B\bar B$.

\par Now if we assume
 the naive factorization the hadronic states of the type
(i) alone completely saturate the parton interference term. Indeed, within this
assumption one would write that the operator $O_2$ in eq. \(6) reduces to
$$\eqalign{&O_2\equiv {1\over N_c}O_1+\tilde O_2,\cr
&\tilde O_2=2(\bar B\Gamma_\mu t^a A)(\bar \beta \Gamma_\mu t^a\alpha ).\cr
}\eqno (8)$$
Here $t^a$ are color
 matrices (${\rm tr} t^at^b={1\over 2}\delta^{ab}$). We neglect
the second term
 ($\tilde O_2$) in eq. \(8) since within the naive factorization
 it has zero projection on the state $(\beta\bar\alpha)+(B\bar \gamma )$.
The correlation function of  $O_1$ and $O_2^+-\tilde O_2^+$
is $1/N_c$ times that of $O_1$ and $O_1^+$ in the naive factorization
 approximation,
and eq. \(7) trivially
 emerges from the factorized duality saturations of two
blocks in Fig.1. There is
 no place for the intermediate states of the type (ii)
if we use the naive factorization technique.
\par The hadronic intermediate state of the type (ii), taken in the
 naive factorization
approximation, just doubles the result. (To see this we rewrite the
operator $O_1$ in the form
 \(8) ) . The contribution of the states of the type
 (iii) is more difficult
  to estimate; one  can argue, however, that it has
the same sign as that of the states (i) and (ii), and, hence,
 only aggravates the
  problem of oversaturation of eq. \(7) within the naive
 factorization.
\par The argument is
 as follows. Although $O_2\ne O_1$,
 and the correlation
  function \(6) is not generally speaking a modulus squared,
in the limit $m_B\rightarrow 0$ each coupling constant enters with its
complex conjugate ( up
 to a symmetric interchange $\beta\leftrightarrow
B$), see Fig. 2,
displaying the hadronic
 saturation of the interference term of the type (iii).
Then
 it is natural to
  expect that at least  in some interval of the B-quark mass
this coincidence in the signs will persist.
\par Thus, we see that the naive factorization is
 incompatible with the asymptotic parton  formulae
plus duality. Of course,
 the typical invariant mass of the hadrons saturating
eq. \(7) is of order
 $\Delta $-large in the usual hadron scale. Therefore, our
analysis does not rule out
the possibility of the
 naive factorization in the specific kinematics considered
in \ref{18}. The statement of the naive factorization in the formulation
of \ref{18} is akin to analogous factorization theorems proven in
 perturbation theory in the formfactor
problem \refto{19}.
\par The considerations
 above are a straightforward kinematic generalization
 of the assertion of \ref{11} . There  a specific (imaginary ) quark
transition has been engineered, with an additional $SU(2)_f$ symmetry
built in. This
 symmetry is further used to argue that a universal suppression
factor $\sim 1/2$
 introduced by hand in the naive factorization prescription
 reconciles the latter with the asymptotic formulae and duality. In the
realistic case
 discussed here there is no additional $SU(2)_f$ to help us to
established the degree of suppression.
\par From the analyses presented above a
whole spectrum of possibilities emerges. The extreme scenarios are as
follows: the
 saturation of eq. \(7) is entirely due to the states (i) with
no contribution
 from (ii) and (iii), or, on the contrary, the saturation is due
to the states
 (ii), and those of the states of the type (i) and (iii) decouple,
in violation of naive factorization.
\par If the
 interference term \(7) is further suppressed by the preasymptotic (
power ) effects (section 3), an additional suppression in the saturation
pattern is necessary.
\head{3. Preasymptotic corrections in the interference term}
\par The basic
 theoretical idea allowing one to estimate preasymptotic effects in the
inclusive nonleptonic widths is
 outlined in \ref{3,4}, and we refer the reader
to these papers for further details.
In order to calculate $\Gamma$'s
 we consider the correlation functions of the
type
$$\Gamma_{\rm hadr}(\Had )={1\over
 m_{\Had} }{\rm Im}<\Had\vert \int d^4xiT\{\L_{\rm eff} (x)
\L_{\rm eff}(0)\}\vert \Had >.\eqno (9)$$
We construct  operator
 product expansion, and then take the imaginary part
of the leading term and those
 giving the power corrections. A subtle point here
is that the short distance
 expansion in eq. \(9) is constructed in Minkowski,
not Euclidean kinematics,
 for obvious reasons ($m_A$ is large). Due to this
fact one may be afraid of
 the anomalous terms in the expansion coefficients
of the semiinfrared nature \refto{20} which could never appear in the
truly Euclidean short distance expansion. As a matter of fact, such terms
occur in individual graphs
 \refto{20} rendering the calculation of coefficients
in front of subleading operators rather unpleasant task. Probably
 for this reason the problem of preasymptotic corrections in the inclusive
widths has been abandoned for many years.
\par Recently it has been demonstrated in \ref{21} that if all imaginary
parts in all graphs are
 summed over these anomalous terms cancel each other and
the computation of the
 coefficients in the short distance expansion can be done
in  straightforward manner, using the standard renormalization group to
include logarithmic corrections to lowest-order results.
\par The effects $\sim m_A^{-3}$ due to the spectator quark are studied
in detail in \ref{3,4} .
 We would like now to discuss preasymptotic corrections
where the spectator quark does not participate.
\par In the study of noninterference terms we consider, e.g.  the
correlation function
$$<\Had \vert\int d^4xT\{O_1(x)O_1^+(0)\}\vert \Had >,\eqno (10)$$
and if we keep only the leading
 in $N_c$ contribution and the next-to-leading, the
factorization in two blocks is valid, Fig.1. If so, the preasymptotic
corrections can be calculated separately for each block.
\par We are interested in
 those corrections which distinguish the nonleptonic
width from that of semileptonic decays (trivial factors like $c_{1,2}
,V_1V_2$
aside, of course). It is
 clear that such corrections are concentrated at the
 upper block on Fig.1. The first power correction is, obviously, due to
 insertion of two soft gluons in this block and reduces to the operator
$G^a_{\mu\nu}G^a_{\mu\nu}$. The corresponding effect has the relative
weight $\sim m_A^{-4}$ (or $\Delta^{-4}$).
\par At the same time, in the interference term the first preasymptotic
correction is much bigger. Indeed, consider the correlation function
$$\int d^4x O_1(x)O^+_2(0).\eqno (11)$$
Let us rewrite $O_2$ as
$$O_2={1\over N_c}O_1+\tilde O_2.\eqno (12)$$
It is clear that the
$(\bar \beta \Gamma^\mu\alpha)$ bracket from $O_1$ correlates with the
$\bar \alpha t^a\Gamma_\rho\beta$ bracket from $\tilde O_2^+$  with the
emission of the soft gluon
 field $G^a_{\alpha\beta}$. The operator emerging
has the form
$$(\bar A\Gamma^\mu t^a\tilde G_{\alpha\mu}A).\eqno (13)$$
This is the same operator as the one that appeared
in \ref{1}, and technically each step of the computation parallels that of
\ref{1}.
\par Alternatively, one could rewrite $O_1$ in the form
$$O_1={1\over N_c}O_2+\tilde
 O_1,\quad \tilde O_1=2(\bar\beta\Gamma^\mu t^a
A)(\bar B\Gamma_\mu t^a\alpha ).\eqno (14)$$
Then the operator \(13) should come out from the correlator of the bracket
$\bar B\Gamma^\mu t^a\alpha$
 in $\tilde O_1$ with the bracket $\bar \alpha\Gamma
_\rho B$ in $O_2^+$. It is easy to check that these two alternative
computations produce one and the same result for the correlation
function \(11).
\par The total $1/N_c$ part of the hadronic width of meson $\Had$ is given by
$$\eqalign{
\Gamma_{\rm hadr}&={G^2_F\over 2}\vert V_1V_2\vert^2{1\over m_{\Had}}{2c_1c_2
\over N_c}
{\rm Im}<\Had \vert \int d^4x iT\{O_1(x)O^+_1(0)\}\vert \Had >\cr
&+{G^2_F\over 2}\vert V_1V_2\vert^2{1\over m_{\Had}} 2c_1c_2
{\rm Im}<\Had \vert \int d^4x
 iT\{O_1(x)\tilde O_2^+(0)\}\vert \Had >.\cr}\eqno (14b)$$
The contribution of the first term is given by the eq. \(7).
Consider now the second term. It is easy to check that the correlation
function of $O_1$ and $\tilde O_2^+$ in eq. \(14b) is equal to
$$\eqalign{&
<\Had \vert \int d^4x iT\{O_1(x)\tilde O_2^+(0)\}\vert \Had >\cr
&=-2i^3
<\Had\vert
 \int d^4x e^{-ipx}(\bar A(0)\Gamma_\nu t^aS_B(0,x)\Gamma_\mu A (0))
{\rm tr}(S_\beta (0,x) \Gamma^\mu  S_\alpha(x,0)\Gamma^\nu t^a)
\vert \Had >.\cr}\eqno (14f)$$
where $p$ is the momentum of the A quark ( coinciding with the momentum of
$\Had$ in the limit considered). Moreover, $S(x,0)$ is the quark propagator
in the external gluon field (taken in the Schwinger gauge $x_\mu A^\mu=0$).
\refto{281}
Here $S(x,0)$ is the propagator of the heavy quark in the external gluon
field  in Schwinger gauge $A^\mu x_\mu =0$ \refto{28} :
$$\eqalign{
S(x,0)&={-im^2\over 4\pi^2}{K_1(m\sqrt{-x^2})\over \sqrt{-x^2}}
-{m^2\hat x\over 4\pi^2x^2}K_2(m\sqrt{-x^2})\cr
&+{G^a_{\rho\lambda}t^a\over
 2^5\pi^2}\{{mK_1(m\sqrt{-x^2})\over \sqrt{-x^2}}
(\hat x\sigma_{\rho\lambda}
+\sigma_{\rho\lambda}\hat x)-2imK_0(m\sqrt{-x^2})
\sigma_{\rho\lambda}\}\cr
&+O(G^2)....\cr}\eqno (14m)$$
Here and below the definition of the field strength $G^{\mu\nu}$
includes the coupling constant $g$.
The mass $m$ is the quark mass.
 It is easy to check that in the limit
$m\rightarrow 0$ we get
$$S(x,0)={1\over
 2\pi^2}{\hat x\over x^4}-{1\over 8\pi^2}{x_\alpha\over x^2}
\tilde G^a_{\alpha\beta}t^a\gamma_\beta\gamma_5,\eqno (14h)$$
i. e. the well-known result for the massless case
(see e.g. the review \refto{281} and references therein ).
Using the expressions for propagators \(14m) , \(14h) we obtain for the
correlation function  in eq. \(14f)
$$\eqalign{&
<\Had \vert \int d^4x iT\{O_1(x)\tilde O_2^+(0)\}\vert \Had >\cr
&=-4 <\Had \vert {\int d^4x e^{-ipx} x^{\alpha}m_B^2K_2(m_B\sqrt{-x^2})
 \bar A\gamma_\beta t^a\tilde G^a_{\alpha\beta}A\over 2\pi^6
(\sqrt{-x^2})^6} \vert \Had >.\cr}\eqno (14l)$$
Here $p^2=m^2_A$. Using eqs. (A1)$-$(A3) it is easy to obtain
that the part of
 hadronic width due to the second term in eq. \(14b) is equal to
$$\Delta \Gamma_{\rm hadr}=-{4G^2_F\vert V_1V_2\vert^2
c_1c_2\over m_{\Had} }
{m^2_A\over 48\pi^3}p^\alpha
<\Had \vert  \bar A\Gamma^\beta\tilde G^a_{\alpha\beta}t^a A\vert \Had >
F_2(m^2_B/m^2_A).\eqno (15)$$
 The function
$F_2$ expresses the effect of the nonvanishing B-quark mass, $F_2(0)=1$,
$$F_2(x)=(1-x)^3.\eqno (16)$$
For $m_B=0$ $F_2=1$.
Calculation of this function is sketched in the Appendix.
\par Note, that there is an important technical difference
between the parton model piece, eq. \(4) and the contribution
of the soft-color exchange \(15). In
 the former case in the light-quark loop
only V times V and A times A terms contribute
, by obvious reasons.
In the latter case the  light quark loop
(see the last term in eq. \(14f)) has all three possible
structures $-$ V times V, A times A, and A times V, and
the last one actually doubles the contribution of (V times V +
A times A). (Here by V we mean a vector-like structure and by A an axial
structure).
\par The matrix element of the new operator in the r.h.s. of eq. \(15)
was discussed in detail in \ref{1}. It can be parametrized in terms of
fundamental
parameter $m^2_{\sigma_H}$ which is expressible through the mass splitting
of the vector and
 pseudoscalar heavy-light mesons in the limit $m_A\rightarrow
\infty$. Using the methods of heavy quark effective theory (HQET)
 (see \ref{37,38}) it is possible to obtain \ref{1}
$$m^2_{\sigma_H}=0.35\quad {\rm GeV}^2\eqno (151)$$
and
$$p^\alpha <\Had\vert\bar A\tilde G^a_{\alpha\mu}t^a\Gamma_\mu A\vert \Had >
=2m_A^2m^2_{\sigma_H}.\eqno (152)$$
Let us emphasise that both terms in eq. \(15) are of same order
 in $N_c$. They are $O(1/N_c)$ relatively to the noninterference
contribution, as was expected.
\par Finally, there
 exists       a change of the hadronic width due to the
additional contribution of the operator
$\bar A\tilde G^a_{\alpha\mu}t^a\gamma_\mu A$ in the first term of eq.
\(14b) that arises in the following way. The first term in eq. \(14b) is
proportional to the matrix element of $\bar b\hat \partial
 b$. In order to use equations
of motion to find this matrix element we have to transform it into
the $ \bar b\hat D b$, where $\hat D$ is the covariant derivative. This leads
to the appearance
 of an additional contribution of the magnetic operator simultaneously in
both  the hadronic  and  the semileptonic widths. However the
corresponding effects are
 relatively small and we shall neglect them below.
\par We now summarize our results:
the leading $\sim N^0_c$ interference term in the total hadronic width
arises in the naive
 parton model due to the operator $\bar A A$ and has the form
$$ \Gamma_{1/N_c}={G^2_F\vert V_1V_2\vert^2m^5_A
\over 96\pi^3}F_1(m^2_B/m^2_A)c_1c_2.\eqno (18)$$
The correction to this result due to the operator
$\bar A\vec \sigma \vec H^at^a A$ (operator
 of eq. \(152) in nonrelativistic
limit relevant to large mass expansion)
is
$$\Delta \Gamma_{\rm gl}=-G^2_F\vert V_1V_2\vert^2
m^3_Am^2_{\sigma_H}{1\over
 6}{1\over \pi^3}
c_1c_2F_2(m^2_B/m^2_A).\eqno (19)$$
The ratio of $\Delta \Gamma$ and $\Gamma_{1/N_c}$ is
$$r={\Delta \Gamma_{\rm gl}\over \Gamma_{1/N_c}}
=-16{m^2_{\sigma H}\over m^2_A}
{F_2(m^2_B/m_A^2)\over F_1(m^2_B/m^2_A)}. \eqno (20)$$

\head{4. Numerical estimates.}
\par  We now apply eqs. \(18) , \(19), \(20)
 to the concrete examples of D and B mesons
. Consider first
the D-meson case. In this case
$$r=-16m^2_{\sigma_H}/m^2_c\sim -3.\eqno (21)$$
Here $m_c\sim 1.35$ GeV is
 the charmed quark mass and we neglect the masses of
light $s,d,u$ quarks.
The $1/N_c$ part of hadronic
 decay width is completely cancelled (and even overcompensated) by gluonic
effects. This leads to an
 enhancement $\sim 2.4$ in the total hadronic width
of D-meson.
\par  Note that annihilation mechanism (that also occurs
due to soft gluon exchange) increases D-meson
hadronic width by 10-20 $\%$ \refto{22} . We also recall that interference
mechanism decreases the hadronic width of $D^+$ by $\sim 2.5$
( if we use the value of the
   c-quark mass assumed above).
Taking into account the annihilation and interference contributions and
the color exchange contribution discussed in this paper we obtain the
following results for the inclusive widths of charmed mesons.
For the case of $D^0$ and $F^+$
$${\Gamma_{\rm hadr}\over \Gamma_{\rm tot}}=0.73,\quad
{\Gamma_{\rm s/l}\over \Gamma_{\rm tot}}=0.27.\eqno (22)$$
Here $\Gamma_{\rm s/l}$ is
 the semileptonic branching ratio. This result is
in qualitative agreement with the current experimental data:
$${\Gamma^{\rm exp}_{\rm hadr}\over \Gamma_{\rm tot}}=0.84,\quad
{\Gamma^{\rm exp}_{\rm s/l}\over \Gamma_{\rm tot}}=0.16.\eqno (23)$$
For comparison we also
 give here the predictions of  parton model
without nonperturbative effects
that are equivalent to naive factorization.
$${\Gamma^{\rm naive}_{\rm hadr}\over \Gamma_{\rm tot}}=0.6,\quad
{\Gamma^{\rm naive}_{\rm s/l}\over \Gamma_{\rm tot}}=0.4.\eqno (24)$$
For the case of $D^+$
we have
$${\Gamma_{\rm hadr}\over \Gamma_{\rm tot}}=0.60,\quad
{\Gamma_{\rm s/l}\over \Gamma_{\rm tot}}=0.40.\eqno (25)$$
 This result is in good agreement with
 the current experimental data:
$${\Gamma^{\rm exp}_{\rm hadr}\over \Gamma_{\rm tot}}=0.62,\quad
{\Gamma^{\rm exp}_{\rm s/l}\over \Gamma_{\rm tot}}=0.38.\eqno (26)$$
Here the color exchange that kills $1/N_c$ suppressed part of naive parton
width compensates the decrease of hadronic width due to interference
mechanism.
\par Of course, one must not take the       agreement between the
theoretical predictions and the experimental data
too literally.
The charmed quark is on the boundary of the kinematical
domain where we can use the preasymptotic expansion,
preasymptotic effects are too large to take them quantatively.
 We expect
the accuracy of our results for $r$ to be of order 1.
Nevertheless, we can definitely argue now, that the nonperturbative
contributions due to soft gluon exchange are important, and lead to the
deviation of the inclusive widths from the
predictions of the naive partonic model  in the
right direction.
\par Our results on inclusive widths of D are in qualitative
 agreement with
our previous calculations of exclusive decay widths \refto{1,22}, where
we found that $1/N_c$ rule works well in most of two-particle exclusive
channels. These channels form $\sim 70\%$ of D-meson total hadronic
widths, and the qualitative
  agreement between the results obtained by summing
exclusive decay rates and the results obtained in the direct calculation
of inclusive width shows the self-consistency of the whole approach.
Note, however, that we  obtained here not simply a cancellation
of a  factorizable $1/N_c$ part due to the soft gluon exchange mechanism
, but an overcompensation by a factor of 3.
On the other hand the overcompensation found in two-particle exclusive
decays of D is relatively small, not bigger than $\sim 1.2-1.5$.
 This difference,
 can be connected, first,
   with the inaccuracy of our results due
to the use of the
preasymptotic expansion on the boundary of the kinematical
domain where it is valid (see above). Second, the inclusive widths
calculated in the previous section  include also
the part of the hadronic width due to the  multi-particle
decays ( see, however, the discussion below).
\par Let us now go to B-mesons. If we neglect
the mass of c-quark compared to the b-quark mass, the
  ratio $r$ in eq. \(20) is
equal to
$$r=-16m^2_{\sigma_H}/m^2_b\sim -0.30.
\eqno (27a)$$
Here $m_b\sim 4.5$ GeV is the $b$-quark mass.
In other words, only $\sim 30
 \%$ of $1/N_c$ suppressed naive hadronic width
of B is cancelled due to soft gluon exchange.
Taking into account the c-quark mass slightly improves the situation.
Indeed, for $m_c=1.35$
 GeV and $m_b=4.5$ GeV we have $x=m^2_c/m^2_b\sim 0.1$,
and $F_1(0.1)=0.44$, while $F_2(0.1)=0.72$. Substituting the values of
$F_1, F_2$ in eq. \(20) we obtain the additional enhancement of $r$ by
$\sim 60\%$:
$$r\sim -0.50.\eqno (27)$$
\par  The
contribution of soft gluons    increases (by $\sim 7.5\%$) the
hadronic width
 of B, leading to the ratio $\Gamma_{\rm hadr}:\Gamma_{\rm s/l}=
0.63:0.37$,
 compared to $0.6:0.4$ ratio in the naive factorization approach
( we neglect here small contribution of $\tau -$lepton in semileptonic
 width).
We see that the color exchange mechanism gives contribution in the right
 direction, increasing
  this ratio. However we are still far from experimental
data:
$$\Gamma_{\rm hadr}:\Gamma_{\rm s/l}=
0.77:0.23.\eqno (301)$$
\par These results are in a qualitative agreement with our results
for two-particle decays exclusive decays of B.
 In \ref{1} we showed that $1/N_c$ rule works
well in the decays $B\rightarrow DP(V)$, where $P(V)$ is the light
pseudoscalar (vector) meson. On the other hand there is only $30\%
-50\%$ cancellation of $1/N_c$ part in $B\rightarrow D^* P(V)$ decays.
This is in agreement with the result in eq. \(27).
\par The theoretical and experimental ratios of total hadronic
and semileptonic widths of $B$ are  still in disagreement,
although our result
 changes the predictions of the naive partonic model in the right direction.
 There is however
  no contradiction in our results for inclusive widths of B
and the experimental data.  Note that contrary to D-meson
case the two-particle exclusive decays form only $\sim 10-20\%$ of hadronic
width of $B$.
 So the main source of missing enhancement of hadronic
width of $B$ must lie in
multi-particle decays.
Unfortunately, the
 study of these decays goes beyond the possibilities of our
approach.
\par Let us consider here possible kinematic regions in the weak decays of
heavy mesons, where
one can observe distinctly different dynamic regimes.
 Recall \refto{1} that the color exchange mechanism discussed in this
paper and in \ref{1} assumes the possibility of expansion of amplitudes
in two parameters.
 The first parameter is $\lambda\sim {\mu \Delta\over Q^2}$.
Here $\mu\sim 300-400$ GeV is the typical transverse momentum of light
quarks and gluons in mesons, $\Delta $ is the energy release and
$Q^2\sim 1$ GeV$^2$ is the typical hadronic scale
\footnote*{Q may have also the meaning of the invariant mass of the
quark pair forming a hadronic state of interest, see e.g. eq. (39) below.}.
 The power corrections become important at this scale
in Euclidean domain \refto{31}.
This parameter arises in the following simple way.
The time needed to regenerate the cloud of
the gluons with characteristic momentum
$Q\ge 1$ GeV around the fast light quarks that are formed in the decay of
the  b quark is \refto{23}
$$t_{\rm reg}\sim {\Delta\over Q^2}.\eqno (28)$$
This time must
 be compared to the characteristic time of strong interactions
inside hadron
$$t_{\rm hadr}\sim {1\over \mu}.\eqno (29)$$
Thus the parameter $\lambda$ can be represented as
$$\lambda ={t_{\rm reg}\over t_{\rm hadr}}.\eqno (30)$$
This makes the meaning of this parameter clear:
If $\lambda\gg 1$, then
 nonperturbative effects are not important. The light
 meson
goes away before it could strongly interact with other quarks
and gluons (outside a light meson)  via
soft gluons. This is nothing else than the  color transparency picture,
advocated in \ref{24} (see also \ref{18}).
\par We would like to
 note here an important difference between our analysis
and that of \ref{18}.
 The authors of \ref{18} implicitly assume that
nonperturbative effects
 become important only at the scale of $Q\sim \Lambda_
{\rm QCD}$.
Consequently the expansion parameter in their Large Energy Effective
theory $(LEET)$
 is ${\Delta\over \Lambda_{\rm QCD}}$. If we make such assumption
and also neglect the difference between $\mu$ and $\Lambda_{\rm QCD}$,
we immediately obtain that our expansion parameter $\lambda$ is the same
as theirs.  This
 however is not justified numerically. Indeed,  from
QCD sum rule analyses we know that
nonperturbative effects can not be neglected already at the scales
$Q^2\sim 1$ GeV$^2$. As a result the color transparency picture
of \ref{24} becomes
relevant only for
 much bigger energy releases than it was assumed in \ref{18}.
Numerically, the expansion parameter $\lambda$ is $\sim 0.4$ for D decays
and $\sim 1$ for $B\rightarrow DP(V)$ decays. By no means one can assume
that in these cases $\lambda\gg 1$.
\par If the parameter $\lambda\le 1$ the color transparency picture
does not work,
and we need to take into account the nonperturbative effects
just in the process of the heavy quark decay.
Under additional conditions (see below)
 this can be done using operator
product expansion in Euclidean domain, as in \ref{1}.
\par The parameter $\lambda$ in eq. \(30) is not the only relevant
kinematical parameter. The calculation in \ref{1} assumes the smallness
of another parameter
$$\lambda '={\mu^2\over Q^2}.\eqno (292)$$
The latter parameter measures the strength of power corrections. Only
when $\lambda '\le 1$ we can use the quark-hadron duality
(or the QCD sum rules) to
 transform the quark amplitudes into the hadronic ones.
The duality works well,
 say, in in the light quark channel with $J^P=1^-$.
However, if two mesons
 formed at first stage in the process of the heavy quark
decay are Regge descendants with high spins \footnote*{They may eventually
decay into a large number of vectors and pseudoscalars.},
the quark-hadron duality
 starts at parametrically large values of energy, which
grow with spin, and is not valid, say, for the b quark exclusive decays.
One can say that the parameter
$\mu$ in eq. \(292) is proportional to spins
(see e.g. \ref{251}
 for detailed discussion of
a breaking of quark-hadron
 duality in this "Regge" domain). The higher power
corrections due to nonperturbative effects become increasingly important
in this kinematical region.
\par We can now divide all possible hadronic decays of heavy flavors into
three kinematic domains.
\par (i) $\lambda\le 1,\lambda'\le 1$. In this case the nonperturbative
effects are important and we can use quark hadron duality and OPE
methods to estimate their strength. This is the kinematic domain
analysed in \ref{1} and in the present paper.
All two body exclusive decays of D mesons (i.e. decays
of D mesons to light pseudoscalars, axial mesons and light vector and
tensor mesons) belong to this domain. For B-mesons  all
decays $B\rightarrow DL$
where $L$ is the light meson lie in this domain
(but rather close to its boundary, though)
. According to the
analysis of \ref{1} and the present paper we expect the approximate
rule of discarding $1/N_c$ corrections to be obeyed in this kinematical
domain due to soft gluon exchange.
\par (ii) $\lambda > 1,\lambda' \le 1$. In this kinematic domain
the color transparency
 picture is valid. Consequently, for the decays in this
kinematic domain
 we expect that the naive factorization a la \ref{18,26} is
valid. However , none of the
decays of the type $B\rightarrow D\pi$ seem to lie in this domain.
The only decays that lie in this kinematic domain are $B\rightarrow$
two light mesons. The analysis of these decays can be, perhaps,
 carried using
the methods of perturbative QCD and factorization theorems ( see e.g.
\ref{27, 28}).
\par (iii) Finally, we have the  domain $\lambda'\ge 1$.
This kinematic domain probably includes most of
multi-particle (exclusive) decays of $B$ (if we treat such exclusive
 multiparticle
  as a tree with few branches). Unfortunately, we can not deal
 with this domain at present basing on QCD alone. Note, however,
 that on average the naive factorization does not take place in this
domain (see section 2 and below).
\par After the analysis of
 possible kinematic domains let us return to B meson
decays. Most of its decays ($\sim 80-90\%$) go through
multi-particle  channels,
whose origin is presumably due to Regge descendants.
 If this is true, it is clear
why we do not obtain the big enhancement of hadronic width $\(301)$.
  Indeed, our
approach is valid for those of the exclusive channels that are in the
domain (i). In practice, however,
 most of the relevant decay modes are in the domain (iii)
which can not be analysed by our methods. If we stick only to the decays
in domain (i), then
 ,  we have good agreement between
our results for inclusive and exclusive decays.
\par Finally, we would like
 to caution against the use of naive factorization
technique and the color transparency in the domain (iii).
The ratio in eq. \(301) can be obtained only if the naive factorization is
invalid in the domain (iii) and the QCD effects in this domain lead to
even stronger deviation from the naive factorization than in domain (i),
that was analysed in \ref{1}.
\head{5. Conclusion.}
\par In this paper we discussed the inclusive widths of heavy mesons.
We found that depending on the values of parameters $\lambda$ and
 $\lambda ^{'}$ given by eqs. \(30) , \(292) there are three kinematical
domains possible in weak hadronic decays:
\par (i) $\lambda\le 1,\lambda'\le 1$. In this case the nonperturbative
effects are important and we can use quark hadron duality and OPE
methods to estimate their strength. This is the kinematic domain
analysed in \ref{1} and in the present paper.
\par (ii) $\lambda > 1,\lambda' \le 1$.
This is the domain of the color transparency.
\par (iii) Finally, we have the  domain $\lambda'\ge 1$.
This is the "Regge" domain.
\par The analysis of the previous paper is valid if
 relevant decay modes lie in the domain (i). This is true for the case
of D-meson decays. For the weak decays of charmed mesons we proved that
color gluon exchange  cancels (and even overcompensates $\sim $ 3 times)
 the $1/N_c$   part in the naive
inclusive hadronic width ( see eqs. \(22)
-\(26) and the discussion thereafter)
 . The results are in agreement with the recent experimental
data that supports a strong overall enhancement of hadronic decay width.
\par For the case of the B-meson only
 a modest part of relevant decay modes
is expected to lie in domain (i). Most of the relevant decay modes lie
in the domain (iii), because a substantial part
of  multi-particle decays that form $80-90\%$ of
B-meson hadronic width go via creation of higher orbital excitations..
We found that the color exchange mechanism of \ref{1} cancels about
$50\%$ of  $1/N_c$ suppressed part of inclusive hadronic width calculated
in naive partonic model. This enhances slightly $(\sim 7.5\%)$
the hadronic width of B-mesons. We also argued that such small suppression
rate is due to the fact that our mechanism is applicable only
for the decays in the domain (i). The decays in this domain
include two body exclusive decays that form  $10-20\%$
 of the total hadronic width
of B and possibly some multi-particle decays.
 We also argued that comparison with experimental data
( see eq. \(301)) shows that even stronger suppression of $1/N_c$
parts of the naive parton model amplitudes might happen for
the decays in the kinematic domain (iii), that includes decays
that go through the creation of higher orbital excitations.
 However our approach is not reliable in this
domain because quark-hadron duality is violated.
\par  One of the authors (B.B.)
is grateful to professor M. S.  Witherell for a discussion on the current
experimental situation with the inclusive rates. We acknowledge
very stimulating remarks of N. Uraltsev who informed us about the
cancellation of semiinfrared "anomalous" terms found in \ref{21} and raised
the question of a new analysis of preasymptotic corrections
in the inclusive rates in light of this finding. One of the
authors (M.S) is grateful to A. Vainshtein for useful discussions in
the process of parallel work on the issue.
\head{Appendix.}
We discuss here the simple way of calculating imaginary part of amplitudes
corresponding to a loop with few massless quarks and one massive.
In this way all relevant phase spaces are automatically taken into
account.
 The technique is based on the spectral representation
for integrals that appear in Fourier transform of
the product of several massless and one massive
quark propagators taken in the $x$ representation.
This representation was established in \ref{28}.
We need only the spectral representation for the structure $p^\alpha$.
We are interested in the function
$$p^\alpha
 G(p^2)={\rm Im} \int e^{-ipx}{K_\nu(m\sqrt{-x^2}\over
 (\sqrt{-x^2})^n}x^\alpha ,\eqno
(A1)$$
appearing in front of the structure $p^\alpha$. Here the MacDonald
function $K_{\nu}$ comes from the Green function of the massive quark.
Then
$$G(p^2)=-{\pi^2\over 2^{n-4}m^\nu}{1\over \Gamma ((n-\nu)/2)}
\sum_{k=0}^{k=(n-\nu )/2-1}C^k_{(n-\nu )/2-1}
 U_{(n+\nu )/2-k}^{n+\nu-4}(p^2)
(-1)^{(n-\nu)/2-1-k}.\eqno (A2)$$
Here
$$U^{2p}_k(p^2)={m^{2(k-1)}\over
 (p-1)!}\int^{p^2}_{m^2}{(p^2-z)^{p-1}\over z^k}dz.
\eqno (A3)$$
In this way we can readily take into account a suppression in the phase
space
due to finite c quark mass simultaneously with calculating the inclusive
widths using eqs. \(6) ,\(14b) .
The relevant cases are: n=8, $\nu$=2$-$ we obtain the function $F_1$;
n=6, $\nu=2$ $-$ we obtain the function $F_2$.
\endpage
\references

\refis{1} B. Blok and M. Shifman, preprint NSF-ITP-92/76.

\refis{6} M. Bauer, B. Stech and M. Wirbel, Z. Phys., C34 (1987) 103.

\refis{7} A. J. Buras, J.-M. Gerard and R. Ruckl, Nucl. Phys., B268
(1986) 16.

\refis{5} M. Shifman, Int. J. Mod. Phys., A3 (1988) 2769.

\refis{18} M. Dugan and B. Grinstein, Phys. Lett., B255 (1991) 583.

\refis{31} M. Shifman, A. Vainshtein and V. Zakharov, Nucl. Phys.,
B147
 (1979) 385, 448.

\refis{251} M. Shifman,
 Yad. Fiz., 36 (1982) 1290; [Sov. J. Nucl. Phys., 36
(1982) 749].

\refis{11} M. Shifman, preprint TPI-MINN-91/46-T, 1992 (submitted
to Nucl.
 Phys. B).

\refis{8} A. J. Buras and J.-M. Gerard, Nucl. Phys., B264 (1986) 371.

\refis{15} G. Altarelli and L. Maiani, Phys. Lett., B52, 351
(1974);\hfill\break
           M. K. Gaillard and B. W. Lee, Phys. Rev. Lett., 33 (1974) 108.

\refis{16} R. Ruckl, Weak decays of heavy flavors,
Habilitationsschrift,
           Munich University,
            1983.

\refis{2} B. Guberina, S. Nussinov, R. Peccei, R. Ruckl,
 Phys. Lett., 89B (1979)
 111;\hfill\break
 Y. Koide, Phys. Rev., D20 (1979) 1739;\hfill\break
 I. Kobayashi
  and N. Yamazaki, Prog. Theor. Phys., 65 (1981) 775;\hfill\break
 N. Bilic, I. Guberina, J. Trampetic, Nucl. Phys., B248
  (1984) 268;\hfill\break
 V. Khose and M. Shifman, Uspekhi Fiz. Nauk, 140 (1983) 3 [Sov. Phys.
 Uspekhi 26 (1983) 387].

\refis{3} M. Voloshin and
 M. Shifman, Yad. Fiz., 41 (1985) 187 [Sov. J. of Nucl. Phys.,
41 (1985) 120].

\refis{4} M. Voloshin and M. Shifman, ZheTP, 91 (1986),1180 [Sov. Phys.,
JETP 64 (1986) 698].

\refis{9} W. Bardeen, A.J.
 Buras, J.M. Gerard, Nucl. Phys., B293 (1987) 787.

\refis{10} W. Bardeen,
 A.J. Buras, J.M. Gerard,  Phys. Lett., B192 (1987) 138.

\refis{13}
I. Bigi, N. Uraltsev and A. Vainshtein, preprint TPI-MINN-92/30-T.

\refis{21} I. Bigi and N. Uraltsev, Phys. Lett., B280 (1992) 271.

\refis{12} Ya. I. Azimov,
 L. Frankfurt, V. Khoze, Pisma v ZheTF, 24 (1974) 373
  [JETP Lett, 24 (1976) 338;\hfill\break
 E. Shuryak, Nucl. Phys., B198 (1982) 83;\hfill\break
  M. Voloshin and M. Shifman , Yad. Fiz, 45 (1987) 463
[Sov. J. of Nucl. Phys., 45 (1987) 292].

\refis{17}  H. Pleitschmann,
 Weak Interactions$-$Formulae, Results, and
Derivatives, Springer-Verlag, 1974.

\refis{19} see e.g. J.C. Collins, D.E. Soper, G. Sterman, in Perturbative
 Chromodynamics, ed. A. Mueller, World Scientific, 1989, and references
therein.

\refis{20} M. Voloshin,
 N. Uraltsev, V. Khose, M. Shifman, Yad. Fiz., 46 (1987)
181 [Sov. J. Nucl. Phys., 46 (1987) 112].

\refis{23} Yu. L. Dokshitser, V. A. Khose and  S. I. Troyan, in Perturbative
 Chromodynamics, ed. A. Mueller, World Scientific, 1989.

\refis{37} E. Eichten and B. Hill, Phys. Lett., B234 (1990)
511;\hfill\break
           H. Georgi, Phys. Lett., B240 (1990) 447.

\refis{38} N. Isgur and M. Wise, Phys. Lett. B232 (1989) 113;
           Phys. Lett., B237 (1990) 527.

\refis{24} J. Bjorken,  preprint SLAC-PUB-5278, (1990), Invited Talk at
Les Recontre de la valle d'Aosta, La Thuille, Italy, (1990).

\refis{14}  M. Shifman, A. Vainshtein and V. Zakharov, Nucl. Phys.,
B120 (1977) 316; ZheTF, 72 (1977) 1275 [JETP, 45 (1977) 670].

\refis{26} A. I. Chernyak and A. Zhitnitsky, Phys. Repts., 112 (1984) 175.

\refis{27} see e.g.  S. Brodsky and G.P. Lepage, in  in Perturbative
 Chromodynamics, ed. A. Mueller, World Scientific, 1989.

\refis{28} V.M. Belyaev and B. Blok, Z. fur Physics, C30 (1986) 151.

\refis{22} B. Blok and M. A. Shifman, Yad. Fiz., 45 (1987), 211, 478,
841 [Sov. Journ. Nucl. Phys. 45 (1987) 135; 301; 522].

\refis{281} V. Novikov, M. Shifman, A. Vainshtein, V. Zakharov,
              Fortschr. Phys., 32 (1984) 585.

\endreferences
\endpage
\centerline{\bf Figure Captions.}
\bigskip
{\bf Fig.1:}
 Imaginary part of such graphs determines $\Gamma (A\rightarrow
B\bar\alpha\beta )$ in the asymptotic regime. For the $c_1^2$ part
 communication
  between two blocks separated by the dotted line appears only
 at the level $1/N_c^2$.\hfill\break
{\bf Fig.2:} Factorization versus duality.
\endpage
\endpaper
\end\bye